\documentstyle[epsf,aps,preprint,floats,tighten]{revtex}

\def\gappeq{\mathrel{\rlap {\raise.5ex\hbox{$>$}}
{\lower.5ex\hbox{$\sim$}}}}
\def\lappeq{\mathrel{\rlap{\raise.5ex\hbox{$<$}}
{\lower.5ex\hbox{$\sim$}}}}
 
\begin{document}

\preprint{\vbox{\hbox{CERN-TH/98-271}
                \hbox{hep-ph/9808401}
                \hbox{August 1998}
                }
}

\title{Nucleosynthesis Bounds in Gauge-Mediated Supersymmetry 
Breaking Theories}

\author{T. Gherghetta\footnote{Email: {\tt tony.gherghetta@cern.ch}},
        G.F. Giudice\footnote{On leave of absence from INFN, 
        Sezione di Padova, Italy.} and
        A. Riotto\footnote{On leave of absence from Oxford University, 
        Department of Theoretical Physics, Oxford, U.K. 
        Email: {\tt riotto@nxth04.cern.ch}}}

\address{Theory Division, CERN, CH-1211 Geneve 23, Switzerland}

\maketitle

\begin{abstract}
In gauge-mediated supersymmetry breaking theories the next-to-lightest
supersymmetric particle can decay during or after the nucleosynthesis epoch.
The decay products such as photons and hadrons can destroy the light element
abundances. Restricting the damage that these decays can do leads to 
constraints on the abundance and lifetime of the NLSP. We  
compute the freezeout abundance of the NLSP by
including all coannhilation thresholds which are particularly important in the
case in which the NLSP is the lightest stau. We find that the upper bound on
the messenger scale can be as stringent as $10^{12}$ GeV when the NLSP is the
lightest neutralino and $10^{13}$ GeV when the NLSP is the lightest stau. Our
findings  disfavour models of gauge mediation where the messenger scale  is
close to the GUT scale or results from balancing renormalisable interactions
with non-renormalisable operators at the Planck scale. When combined with the
requirement of no gravitino overabundance, our bound implies that the 
reheating temperature after inflation must be less than $10^7$ GeV. 
\end{abstract}

\newpage

{\bf 1.} Gauge-mediated supersymmetry breaking theories provide an 
interesting alternative to the usual gravity-mediated scenarios 
in transmitting supersymmetry breaking effects to
the low energy world \cite{din} (for a review, see Ref.~\cite{gr}). 
Among the most attractive features of 
these theories is the natural suppression of flavour-violating interactions
which is guaranteed by the gauge symmetry. In addition low-energy 
phenomenology is now governed by the fact that the gravitino is the 
lightest supersymmetric particle and this may lead to interesting collider 
signals.

The next-to-lightest supersymmetric particle (NLSP) also plays an 
important role in low-energy phenomenology. Since the soft mass 
of a supersymmetric particle is determined by its gauge quantum numbers,
the NLSP will be either a neutralino $N_1$ or the lightest, mainly right-handed
stau ${\tilde\tau}_1$, depending on the choice of parameters (the possibility
of a sneutrino NLSP is now marginal). The relevant parameters which define
the gauge-mediated model are the messenger index $N$ (twice the sum of the
Dynkin indices of the messenger gauge representations), and the supersymmetric
and supersymmetry-breaking messenger mass parameters $M$ and $F$.
We also distinguish the supersymmetry breaking scale $F$ felt by the messenger
fields and the fundamental scale of supersymmetry breaking $F_0$ which
determines the gravitino mass and couplings by defining $k\equiv F/F_0\leq 1$.
In our analysis, we will trade $F$ for the NLSP mass which is a parameter
of more direct physical meaning. The Higgs mass parameters $\mu$ and $B_\mu$
are new independent inputs, which can be 
determined by imposing electroweak symmetry breaking, and therefore
be fixed in terms of $\tan \beta$ and the algebraic sign of $\mu$. As a 
result, the lightest neutralino turns out to be mainly $B$-ino. More
details on the definition of the parameters can be found in Ref.~\cite{gr}.

The NLSP is not stable and eventually will decay 
into the gravitino. If these decays occur 
during the nucleosynthesis epoch the light element abundances can be 
drastically altered~\cite{limnuc,rs,dim,ell,kaw,ddrg,hkkm}. 
For example, if the NLSP is a neutralino the dominant 
decay mode produces a photon which affects the primordial
$^4$He abundance.
On the other hand hadronic decays can also prove dangerous for the
nucleosynthesis products. In the case of the neutralino NLSP, the photon
can hadronise or, in the case of a stau NLSP, the semileptonic decay of a 
tau lepton can produce hadronic showers which leads to energetic nucleons.
Even though nucleosynthesis may be over, these energetic nucleons
destroy $^4$He or synthesise $^3$He and tritium leading to
the overproduction of the light elements. 
Alternatively if the NLSP actually decays hadronically during 
nucleosynthesis these nucleons will instead 
establish thermal equilibrium with the surrounding plasma 
by colliding with the ambient protons and $^4$He leading to the eventual
increase of the $n/p$ ratio. This results in a greater abundance of 
$^4$He, $^3$He and deuterium D.

In order to avoid the destructive effects on the nucleosynthesis 
products the lifetime of the NLSP must be restricted so as to decay 
sufficiently well before it can interfere with the nucleosynthesis 
products or if it decays during nucleosynthesis that the enhanced 
light element abundances are consistent with astrophysical observations. 
In addition the abundance of the NLSP at the time of decay will also
be important. Since the abundance and lifetime are 
related to the messenger scale $M$, an upper bound can be placed
on the messenger scale (for a typical set of the other 
gauge-mediated parameters), depending on whether the neutralino or stau is
the NLSP. These bounds will be shown to be fairly generic
for most of the parameter space of gauge-mediated theories.

{\bf 2.} The damaging effects of the NLSP decay products during the 
nucleosynthesis epoch constrains the abundance and lifetime 
of the decaying particle. In order to obtain a bound on the NLSP
lifetime a detailed calculation of the NLSP abundance at the time of decay
must be performed. This amounts to calculating the NLSP abundance
at the time of freeze out when the NLSP is no longer in chemical 
equilibrium. 

We will consider the two separate cases of a neutralino and stau NLSP with
mass $m_{\rm NLSP}$.
For moderate values of $\tan \beta$, ${\tilde\tau}_1$ is lighter 
than the neutralino whenever 
\begin{equation}
N>\frac{66}{5(13\xi -2)},~~~~\xi\equiv \frac{\alpha_1^2(m_{\rm NLSP})}
{\alpha_1^2(M)}=\left[ 1+\frac{22}{4\pi}\alpha_1(m_{\rm NLSP})\ln 
\frac{m_{\rm NLSP}}{M}\right]^2~.
\label{limn}
\end{equation}
For large $\tan\beta$ this region becomes slightly larger because of the
stau left-right mixing.
In order to determine the
NLSP abundance we consider all relevant channels which
change the number of NLSP's. When the NLSP is a neutralino
the most relevant annihilation channels are those consisting of fermions
and gauge bosons as depicted in Table~\ref{annch}. The annihilation channels
into Higgs bosons are suppressed because the lightest neutralino is mainly
$B$-ino. For the same reason, 
the neutralino NLSP generically is not degenerate in mass with other particles,
so it will not be important to 
consider coannihilations for this case. We can also neglect annihilation
channels which involve gravitino vertices since these lead to scattering 
amplitudes which are suppressed by a factor $m_{\rm NLSP}^2/F_0$. Since 
we are interested in NLSP lifetimes of the order of the nucleosynthesis 
timescale, this suppression makes all gravitino-emission processes negligible.

On the other hand when the NLSP is the stau, the lightest smuon and 
selectron have a small mass difference with the NLSP:
\begin{equation}
\frac{m_{ \tilde{\mu}_1,\tilde{e}_1 }-m_{ \tilde{\tau}_1 } } 
{ m_{ \tilde{\tau}_1 } }\simeq
\frac{m_\tau^2}{2m_{ \tilde{\ell}_R }^2}\left[ \frac{(\mu \tan\beta -A_\tau)^2}
{m_{ \tilde{\ell}_L }^2-m_{ \tilde{\ell}_R }^2}-1\right] ~.
\label{massd}
\end{equation}
Here $A_\tau$ is the supersymmetry-breaking trilinear term and 
$m_{ \tilde{\ell}_{L,R} }$ are the flavour-independent
contributions to the left and right slepton masses
(including the $D$-term contribution). Since in gauge-mediated theories
$\mu^2>m_{ \tilde{\ell}_L}^2$, the mass difference in Eq.~(\ref{massd})
is always positive. Because of the approximate mass degeneracy among the 
sleptons, the calculation of the stau relic abundance must include all
coannihilation processes listed in Table~\ref{annch}. We also need to
compute the smuon and selectron density at the decoupling time, since
these ``co-NLSP's'' are also responsible for producing
damaging decay products.

The NLSP abundance is determined by considering the evolution of 
the number density $n_i$ of particle $i$ which is governed by the 
Boltzmann equation. In the presence of coannihilations the Boltzmann 
equation can be written as~\cite{eg}
\begin{equation}
   {dn\over dt} = -3Hn-\langle\sigma_{\rm eff} v \rangle(n^2-n_{\rm eq}^2)
\end{equation}
where $H$ is the Hubble expansion parameter, 
$n=\sum_i n_i$ and the thermal average of the
effective cross section is defined as
\begin{equation}
\label{thav}
  \langle\sigma_{\rm eff} v \rangle =
    \sum_{ij}\langle\sigma_{ij} v_{ij}\rangle {n_i^{\rm eq}\over n^{\rm eq}}
    {n_j^{\rm eq}\over n^{\rm eq}}.
\end{equation}
The individual cross sections $\sigma_{ij}$ include the processes 
listed in Table~\ref{annch}, $v_{ij}$ is the relative velocity and
$n^{\rm eq}$ is the equilibrium number density. 
\begin{table}
\caption{Final state annihilation channels including coannihilations in
parentheses.}
\label{annch}
\begin{tabular}{cc}
initial state & final state \\
\hline
$N_1\,N_1$ & $ZZ, W^+W^-, {\bar f}f$ \\
${\tilde\tau}_1^+ {\tilde\tau}_1^-, ({\tilde\mu}_1^+ {\tilde\mu}_1^-, 
{\tilde e}_1^+ {\tilde e}_1^-)$ 
& $Z Z, W^+W^-, \gamma\gamma, Z \gamma , h h,\gamma h, Z h, {\bar f}f$ \\
${\tilde\tau}_1^\pm {\tilde\tau}_1^\pm, 
({\tilde\mu}_1^\pm {\tilde\mu}_1^\pm, {\tilde e}_1^\pm {\tilde e}_1^\pm)$
& $\tau^\pm \tau^\pm, (\mu^\pm \mu^\pm, e^\pm e^\pm)$ \\
$({\tilde\tau}_1^\pm {\tilde\mu}_1^\pm, 
{\tilde\tau}_1^\pm {\tilde e}_1^\pm, {\tilde\mu}_1^\pm {\tilde e}_1^\pm)$ 
& $(\tau^\pm \mu^\pm, \tau^\pm e^\pm, \mu^\pm e^\pm)$ \\
$({\tilde\tau}_1^\pm {\tilde\mu}_1^\mp, {\tilde\tau}_1^\pm {\tilde e}_1^\mp,
{\tilde\mu}_1^\pm {\tilde e}_1^\mp)$ & 
$(\tau^\pm \mu^\mp, {\bar\nu}_\tau \nu_\mu, \nu_\tau {\bar \nu}_{\mu}),
(\tau^\pm e^\mp, {\bar\nu}_\tau \nu_e, \nu_\tau {\bar \nu}_e),
(\mu^\pm e^\mp, {\bar\nu}_\mu \nu_e, \nu_\mu {\bar \nu}_e)$ \\
\end{tabular}
\end{table}

The cross sections for all annihilation channels are numerically
evaluated using the ${\tt CompHEP}$ software package~\cite{comphep}. 
After performing the thermal average and including all relevant coannihilation 
thresholds, the ratio $Y^{\rm eq}$ of the equilibrium number 
density $n^{\rm eq}$ to the entropy density $s$ is given by
\begin{equation}
\label{yeq}
  Y_{\rm eq}(T)\equiv {n^{\rm eq}\over s}={45 x^2\over 4 \pi^4 g_{\rm eff}(T)}
  \sum_i g_i {m_i^2\over m_{\rm NLSP}^2} K_2\left(x {m_i \over
  m_{\rm NLSP}}\right),
\end{equation}
where $x=m_{\rm NLSP}/T$, $g_i$ is the number of internal degrees
of freedom and $g_{\rm eff}=81$. The NLSP abundance 
$Y_{\rm NLSP}$ at the time
of nucleosynthesis is given by $Y_{\rm NLSP}\equiv Y_{\rm eq}(T_F)$, 
where $T_F$ is the
freezeout temperature. The freezeout temperature is determined from the 
condition $n_{{\rm eq}}\langle \sigma_{\rm eff} v \rangle =H(T)$.

\begin{figure}[ht]
\centerline{ \epsfxsize 5.8 truein \epsfbox {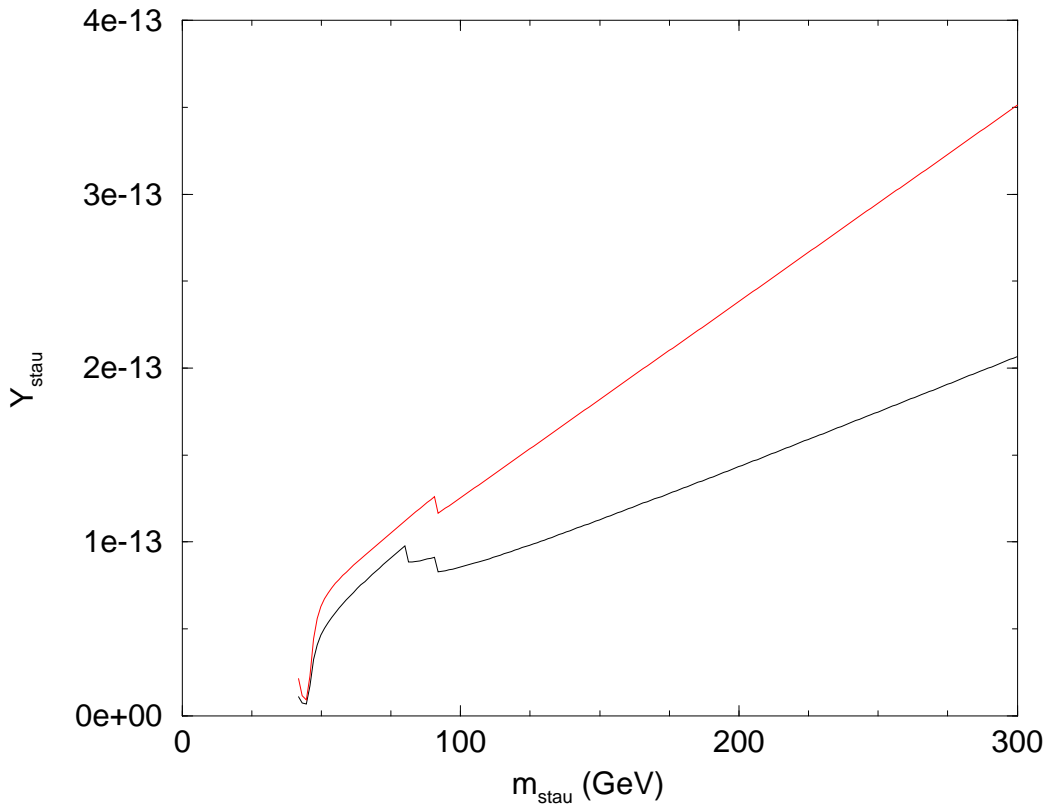}}
\caption{\it Comparison of the stau abundance with and without
coannihilation (lower and upper curves, respectively). 
The gauge-mediated parameters are
$M=10^{13}$ {\rm GeV}, $N=12$, $\tan\beta=1.1$ and ${\rm sgn}\mu=1$.}
\label{stauabundfig}
\end{figure}

Thus in the case of the 
neutralino NLSP the abundance $Y_{\rm NLSP}$ is just simply 
$Y_{N_1}$ while for the stau NLSP it will be 
$Y_{\rm NLSP}=Y_{\tilde\tau}+Y_{\tilde\mu}
+Y_{\tilde e}$ which is the sum of all the ``co-NLSP'' abundances. 
The effect of including the coannihilating channels between the
stau and the ``co-NLSP's'' changes the pure stau abundance by up to
$\sim 50\%$. 
This can be seen in Fig.~\ref{stauabundfig} where the stau abundance 
with and without coannihilations is shown. Notice the important reduction
of the relic abundance for values of $m_{{\tilde \tau}_1}$ close to half
the Higgs mass, caused by the resonant annihilation channel.
It should be noted, however that given the present LEP bound on 
stable ${\tilde \tau}_1$, this possibility is no longer realistic 
for the stau.

{\bf 3.} Let us now consider the effect of the NLSP decaying during 
the nucleosynthesis epoch. Suppose first that the NLSP is a neutralino. 
The dominant decay 
mode of the neutralino is into a photon and gravitino with a decay rate 
given by~\cite{gr,ambros,dtw}
\begin{equation}
   \Gamma(N_1\rightarrow \gamma {\tilde G}) = {k^2\kappa_\gamma m_{N_1}^5
     \over 16\pi F^2}
\end{equation}
where $\kappa_\gamma=|N_{11}\cos\theta_W+N_{12}\sin\theta_W|^2$ and 
$N_{11,12}$ are the NLSP gaugino components in standard notation. There are
also other decay modes which involve the $Z$ or a Higgs boson, but these modes
are suppressed by a $\beta^8$ phase-space factor. The photon of the dominant
decay mode has a dramatic effect on the light element 
abundance\footnote{Notice that the gravitinos released in the NLSP decays 
do not thermalise and their present contribution 
to the energy density of the Universe is  negligible.}. In the 
radiation dominated thermal background high energy photons initiate 
electromagnetic cascades and create many low-energy photons capable of 
photodissociating the light elements~\cite{limnuc}. 
These photodissociation effects caused by the photon of the 
decaying neutralino can be parametrised by a ``damage'' factor
\begin{equation}
\label{dgamma}
    d_\gamma\equiv m_{N_1} {Y_{N_1}\over \eta}
\end{equation}
where $\eta$ is the baryon-to-photon number density ratio. This 
quantity $d_\gamma$ is constrained by astrophysical observations 
of light elements. Recently new observations of the 
$D$ and $^4$He primordial abundance have led to a reanalysis of 
the constraints on the abundance and lifetime of long-lived 
particles~\cite{hkkm}.
In particular, controversy
in the measurement of primordial deuterium have led to 
two different values for the deuterium abundance $X_D$ 
(normalised to hydrogen). In Ref.~\cite{rh} a high value of the 
deuterium abundance $X_D=(1.9\pm0.5)\times 10^{-4}$ is quoted
while in Ref.~\cite{bt} a low value $X_D=(3.39\pm0.25)\times 10^{-5}$
is obtained. 

\begin{figure}[ht]
\centerline{ \epsfxsize 5.8 truein \epsfbox {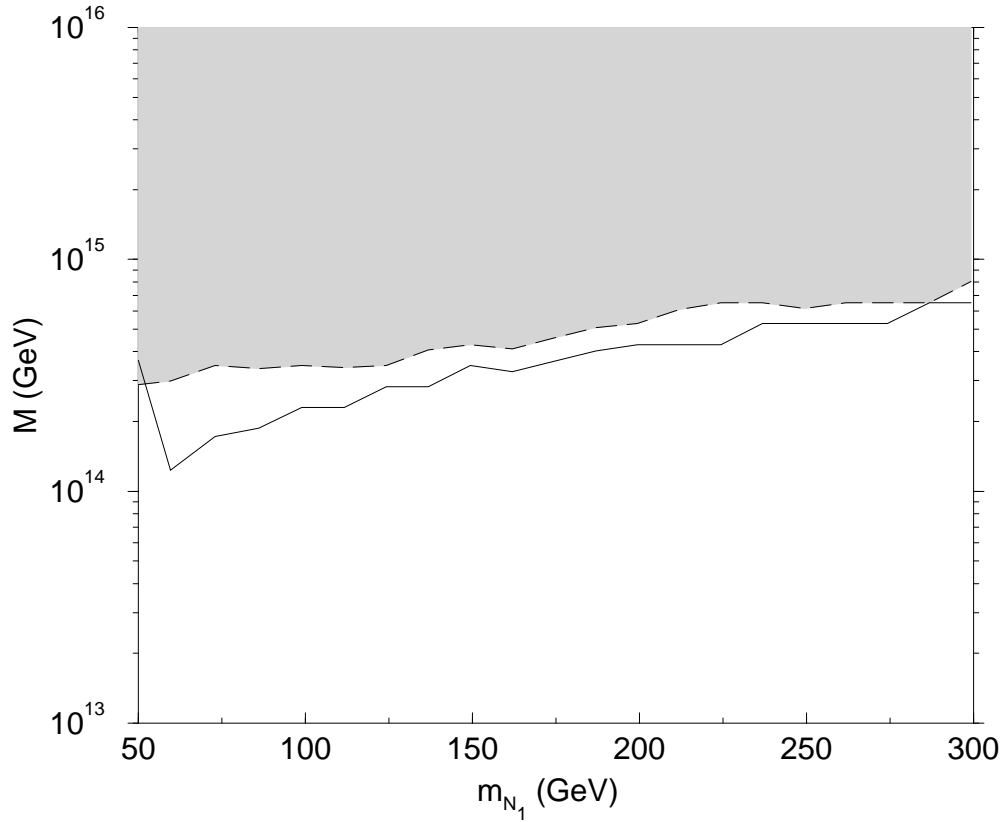}}
\caption{\it Upper bound on the messenger scale $M$ as a function
of the neutralino NLSP mass $m_{N_1}$ from photodissociations
where $k=1$, $N=2$,
$\tan\beta=2$ and ${\rm sgn}\mu=1$. Both bounds use a 
$^4${\rm He} abundance $Y_p(^4{\rm He})= 0.234$ 
but the dashed line uses a low value of the
deuterium abundance while the solid line uses the high value of the 
deuterium abundance (see text).}
\label{dgfig}
\end{figure}

The neutralino abundance $Y_{N_1}$ depends on the details of the 
sparticle spectrum and can be calculated for a particular
choice of the parameters $N$, $M$, $F$, $\tan\beta$ and ${\rm sgn}\mu$.
Using the constraints in 
Ref.~\cite{hkkm} we show in Fig.~\ref{dgfig} the bound on the messenger 
scale for a representative neutralino NLSP 
scenario in which $N=2$, $\tan\beta=2$ and ${\rm sgn}\mu=1$, for both
high and low values of the measured $X_D$. We have also assumed that 
$k=1$ and the limiting
value of $M$ grows approximately linearly with $k$. Notice that the high
deuterium value gives a slightly more stringent bound than the low
deuterium value, because for lifetimes less than $10^6$ seconds deuterium
is effectively photodissociated.

The bound will change slightly for different
values of the gauge-mediated supersymmetry breaking parameters, but 
typically for the neutralino NLSP scenario lifetimes greater than 
$10^5$ seconds are ruled out by photodissociations.

Note that photodissociations are only important after 
nucleosynthesis is over (which corresponds to times later than 
$10^4$ seconds). This is because during earlier times when thermal 
photons are more energetic and numerous, photon-photon interactions are 
much more probable than photon-nucleus interactions. Consequently 
stronger constraints can only be obtained if we consider processes 
that affect nucleosynthesis at times earlier than $10^4$ seconds.

However before discussing the relevant processes for times earlier than
$10^4$ seconds it is possible to obtain stronger constraints than those
from photodissociations by considering hadronic decays of the 
NLSP~\cite{ddrg}.
If the NLSP decays hadronically during times $\gtrsim 10^4$ seconds
then the photodissociation bound needs to be reconsidered since while
D is destroyed by photodissociation it can be produced by the hadronic 
showers. In fact even if the NLSP decays exclusively into photons it is 
possible that hadronic showers will be generated. The main effect 
of hadronic showers is not only to increase the abundance of D but also
the other light nuclei $^3{\rm He}, ^6{\rm Li}$ and $^7{\rm Li}$.
In particular D and $^3$He arise mainly from ``hadrodestruction'' 
processes, which occur when an energetic nucleon breaks up the 
ambient $^4$He nucleus into D and $^3$He. The other light nuclei,
$^6{\rm Li}$ and $^7{\rm Li}$ arise from ``hadrosynthesis'' of
$^3$He, T or $^4$He. More complete details can be found in Ref.~\cite{dim}.
Again the ``damage'' of the hadronic decays on the primordial nuclear
abundances can be parametrised as
\begin{equation}
\label{dB}
    d_B\equiv d_\gamma B_h^\ast
\end{equation}
where $B_h^\ast=(\nu_B/5)B_h {\cal F}$ is an effective baryonic branching 
ratio which depends on the true baryonic branching ratio $B_h$, the baryonic
multiplicity $\nu_B$ and a factor ${\cal F}$ representing the
dependence of the yields on the energy of the primary shower baryons.

One of the contributions to the neutralino
hadronic branching ratio comes
from the decay $N_1\rightarrow Z\tilde{G}$
\begin{equation}
   B_h\simeq 0.7 \tan^2\theta_W\left(1-{m_Z^2\over m_{N_1}^2}\right)^4
\end{equation}
where the hadronic branching ratio of the $Z$-boson is 0.7, and we have
assumed a pure $B$-ino. Although this contribution, and the
analogous one from Higgs decay, can be forbidden by
phase space, there is always at least a contribution to $B_h$ of order 
$\alpha$ from photon conversion into a $q\bar q$ pair. The explicit 
expression for $B_h$ can be obtained from Eq.~(13) of Ref.~\cite{bagger}.
Using the corresponding value of $d_B$,
one finds that the overproduction of $^7$Li constrains the lifetime 
of the neutralino to be shorter than $10^4$ seconds. This is also true 
if the NLSP is the stau. The stau decays predominantly into a tau 
and gravitino and since the tau has a large hadronic branching ratio, 
the corresponding value of $d_B$
is much larger than that for the neutralino. Consequently the lifetime of
the stau must also be less than $10^4$ seconds if the overproduction
of $^7$Li is to be avoided~\cite{ddrg}.

It is clear that in order to obtain precise constraints on the abundance
and lifetime of the NLSP we need to consider processes occuring at times
earlier than $10^4$ seconds. At these times the main decay products
that interfere with nucleosynthesis are hadrons. Hadronic showers induce 
interconversions between the ambient protons and neutrons thus changing 
the equilibrium $n/p$ ratio~\cite{rs}. In particular during the lifetime
interval $\tau_{\rm NLSP}\sim 1 -100$ seconds the overall effect of the
hadronic decays is to convert protons into neutrons. The additional neutrons
that are produced are all synthesised into $^4$He and thus hadronic decays 
during this time interval are constrained by the observational upper bound
on the primordial helium abundance $Y_p(^4{\rm He})$.

Eventually the neutron fraction falls to zero because all neutrons 
are contained in the $^4$He nuclei and the remaining neutrons 
created by the NLSP decay increase the deuterium D abundance. Furthermore 
for $\tau_{\rm NLSP}\sim 100-1000$ seconds the $^3$He abundance is also 
increased by D-D burning. After $\tau_{\rm NLSP}\gtrsim 10^4$ seconds
all the neutrons arising from NLSP decay will themselves decay before forming
D. Thus in the interval $10^2-10^4$ seconds the appropriate 
constraint on hadronic decays arises from the observational bounds
on $({\rm D}+^3{\rm He})/{\rm H}$.

The overall effect of these hadronic decays has been considered 
in Ref.~\cite{rs}, where the constraints on the abundance and 
lifetime of the decaying particle are parametrised by
\begin{equation}
    f={N_{jet} B_h\over 2} {\langle n(E_{jet})\rangle \over
      \langle n(33\,{\rm GeV})\rangle}
\end{equation}
where $N_{jet}$ is the number of jets, $B_h$ is the hadronic branching ratio
and $\langle n(E_{jet})\rangle$ is the average charge multiplicity for 
a jet with energy $E_{jet}$ and is given by
\begin{equation}
    \langle n(E_{jet}) \rangle =1+0.0135\,
    \exp\left[ 1.9\sqrt{ 2\ln\left({E_{jet}\over 0.15~ {\rm GeV}}\right)
    } \right].
\end{equation}
Since at the parton level the neutralino NLSP decays into
three particles we will assume that $E_{jet}=m_{\rm NLSP}/3$ and $N_{jet}$
is the number of quarks at the parton level. In the case of the stau
in which the tau decays semi-leptonically we assume $E_{jet}=m_{\rm NLSP}/6$.

The most stringent constraints on the NLSP abundance and 
lifetime in the interval $10^2-10^4$ seconds come from the 
primordial deuterium abundance. Previous constraints on $Y_{\rm NLSP}\,f$ 
and the lifetime were obtained for $X_D<10^{-4}$ and 
$\eta=3\times 10^{-10}, 10^{-9}$~\cite{rs}. 
In order to use the more recent measurements, we rescale the previous 
constraints for the new values  
$X_D=(1.9\pm0.5)\times 10^{-4}$ or
$X_D=(3.39\pm0.25)\times 10^{-5}$. 
This rescaling can be done analytically by using 
the change in the 
deuterium abundance~\cite{rs}
\begin{equation}
\Delta X_D =
\frac{\Delta Y_{\rm NLSP}}{Y_H}\epsilon_D a_n,
\end{equation}
where $Y_H$ is the hydrogen density, $\epsilon_D$ is the fraction of injected
neutrons that end up in deuterium and $a_n$ is the number of $n\bar n$ pairs
per NLSP decay. This equation is valid as long as $X_D\ll \epsilon_D$, which
conveniently holds whenever the deuterium constraint is important.
Since the limit comes from the
overproduction of deuterium and the standard $X_D$ prediction decreases
with $\eta$, we make the most conservative choice of $\eta =6\times 10^{-10}$,
which is the largest value compatible with $^4$He and $^6$Li abundances.

When the high value of the deuterium abundance is used, 
one obtains no constraints in the region
$10^2-10^4$ seconds. All values of the NLSP abundance and lifetime 
are consistent with high $X_D$ value and consequently there is no 
improvement on the lifetime upper bound of $10^4$ seconds obtained from the 
overabundance of $^7$Li.

\begin{figure}[ht]
\centerline{ \epsfxsize 5.8 truein \epsfbox {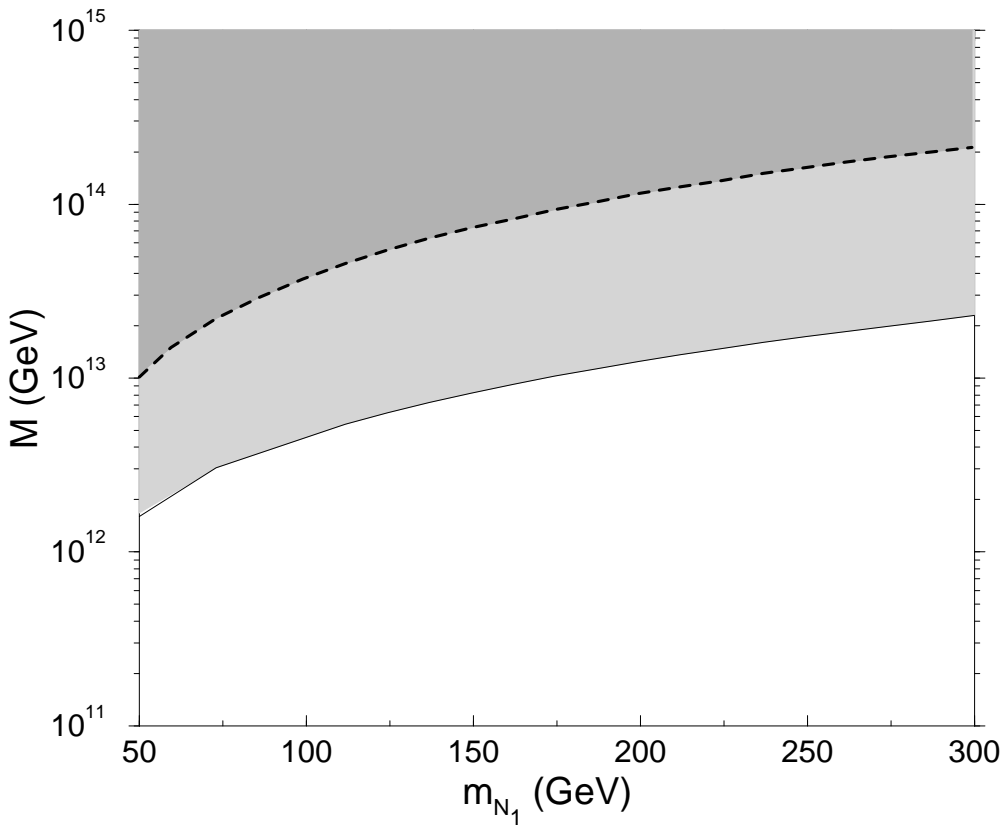}}
\caption{\it Upper bound on the messenger scale $M$ as a function
of the neutralino NLSP mass $m_{N_1}$ from hadronic decays
where $k=1$, $N=2$, $\tan\beta=2$ and ${\rm sgn}\mu=1$. 
The solid (dashed) line corresponds to the bound assuming a low (high)
deuterium measurement. More specifically the
dashed line represents the $10^4$ second lifetime contour which arises from
the $^7$Li overabundance.}
\label{dBneutfig}
\end{figure}

On the other hand stringent constraints in the interval $10^2-10^4$ seconds
are obtained when the low value of the deuterium abundance is used. Let us 
consider the two NLSP scenarios separately. First, when the NLSP is the 
neutralino the scaling parameter $f$ is determined using 
$N_{jet}=2$ and $B_h=10^{-2}$. Rescaling the solid curve of Fig.~4 in 
Ref.~\cite{rs} enables one to determine the constraint arising from the low
value of $X_D$ and $\eta=6\times 10^{-10}$.
Thus calculating the value for the neutralino number 
density for a generic set of gauge-mediated supersymmetry breaking 
parameters leads to constraints on the abundance and 
lifetime of the neutralino which 
can be expressed as a bound on the messenger scale $M$, see 
Fig.~\ref{dBneutfig}. The bound on $M
$
does not significantly depend on the value
of $\tan\beta$, but grows approximately linearly with $N$.

Let us now discuss 
the case of the stau
NLSP. At the decoupling time, the abundances of smuons and selectrons are
comparable to the stau abundance.
The cosmological fate of the frozen-out smuons and selectrons depends on
the mass
difference in  Eq.~(\ref{massd}). If $\tan \beta$ is large, 
then $m_{ \tilde{\ell}_1 }-
m_{ \tilde{\tau}_1 }>m_\tau - m_\ell$ ($\ell =e,\mu$) and the decay
$\tilde{\ell}_1\to \tilde{\tau}_1 \tau \ell$ is kinematically open. The
value of $\tan\beta$ for which this
transition occurs depends on the parameter choice, but it is typically of
order 10, as it can be estimated from Eq.~(\ref{massd}). If this mode
is accessible, 
it dominates over the two-body decay into gravitino.
This can be simply understood from the decay rate expression~\cite{ambm}
in the limit
of vanishing lepton masses and small slepton mass difference,
\begin{equation}
\Gamma (\tilde{\ell}_1^-\to \tilde{\tau}_1^- \tau^+ \ell^-)
\simeq \frac{4G_F^2}{15\pi^3} \tan^4\theta_W \frac{M_W^4}{m_{N_1}^4}
(m_{ \tilde{\ell}_1 }-
m_{ \tilde{\tau}_1 })^5,
\end{equation}
\begin{equation}
\Gamma (\tilde{\ell}_1^-\to \tilde{\tau}_1^+ \tau^- \ell^-)
\simeq \frac{4G_F^2}{15\pi^3} \tan^4\theta_W \frac{M_W^4}{m_{N_1}^2
m_{ \tilde{\ell}_1 }^2}
(m_{ \tilde{\ell}_1 }-
m_{ \tilde{\tau}_1 })^5.
\end{equation}
Here $m_{N_1}$ is the mass of the $B$-ino which mediates the decay, and
is assumed to be much larger than $m_{ \tilde{\ell}_1 }$.
In this case, smuons and selectrons decay soon after decoupling, with
each one producing
a stau in the final state. Therefore, the relevant stau abundance at
the nucleosynthesis epoch is determined by the sum over the three slepton 
abundances at the decoupling time. 
The stau hadronic branching fraction comes from the semileptonic tau 
decay, and it is given by $B_h\simeq 0.65$. 

For smaller values of $\tan\beta$ the neutralino-mediated three-body process
is forbidden, and the rates of the two competing slepton decay modes are
\begin{equation}
\Gamma (\tilde\ell_1 \to \ell \tilde G)=\frac{k^2m_{\tilde\ell_1}^5}{16\pi
F^2}
\label{dgra}
\end{equation}
\begin{equation}
\Gamma (\tilde{\ell}_1^-\to \tilde{\tau}_1^- \nu_\ell \bar \nu_\tau )
\simeq \frac{G_F^2}{15\pi^3} \frac{M_W^4}{m_{\chi^+}^4}
(m_{ \tilde{\ell}_1 }-
m_{ \tilde{\tau}_1 })^5 \sin^2\theta_{\tilde \tau}\sin^2\theta_{\tilde \ell}.
\label{dneu}
\end{equation}
Here $m_{\chi^+}$ is the chargino (gaugino) mass which mediates the decay,
and $\theta_{\tilde \tau}$, $\theta_{\tilde \ell}$ are the left-right mixing
angles in the slepton system. The process in Eq.~(\ref{dgra}) is suppressed
by the large value of $F$ required in our study of nucleosynthesis and
the process in Eq.~(\ref{dneu}) is suppressed by the small mixing angles
proportional to the corresponding lepton masses. It turns out that in the
small $\tan\beta$ regime and for $m_{ \tilde{\ell}_1}\sim 100$ GeV, the two
rates are comparable when $\sqrt{F}\sim 10^9$ GeV, which is just the value 
of $\sqrt{F}$
necessary to have a stau decay during nucleosynthesis. Therefore, depending
on the particular choice of the gauge mediation parameters, we can
encounter different situations. The first possibility is that 
$\Gamma (\tilde\ell_1 \to \ell \tilde G)>\Gamma (\tilde{\ell}_1^-\to 
\tilde{\tau}_1^- \nu_\ell \bar \nu_\tau )$, in which all light sleptons
decay directly into gravitinos around the same time. In this case smuons
and selectrons do not significantly contribute to the effective hadronic
branching fraction, because real electrons and muons cannot
decay into hadrons. As $\tan\beta$ is increased and $m_{ \tilde{\ell}_1}$ is
decreased, we go first to a regime in which $\Gamma (\tilde{\mu}_1^-\to 
\tilde{\tau}_1^- \nu_\mu \bar \nu_\tau )>
\Gamma (\tilde\ell_1 \to \ell \tilde G)>
\Gamma (\tilde{e}_1^-\to 
\tilde{\tau}_1^- \nu_e \bar \nu_\tau )$ and then in a regime in which the
three-body decays dominate for both $\tilde{\mu}_1$ and $\tilde{e}_1$.
The two regimes are possible because the decay process in Eq.~(\ref{dneu})
depends on the lepton mass through the left-right mixing angle. In the
first regime, only the smuon contributes to the effective hadronic
branching fraction, while in the second one both the 
smuon and selectron contribute.

\begin{figure}[ht]
\centerline{ \epsfxsize 5.8 truein \epsfbox {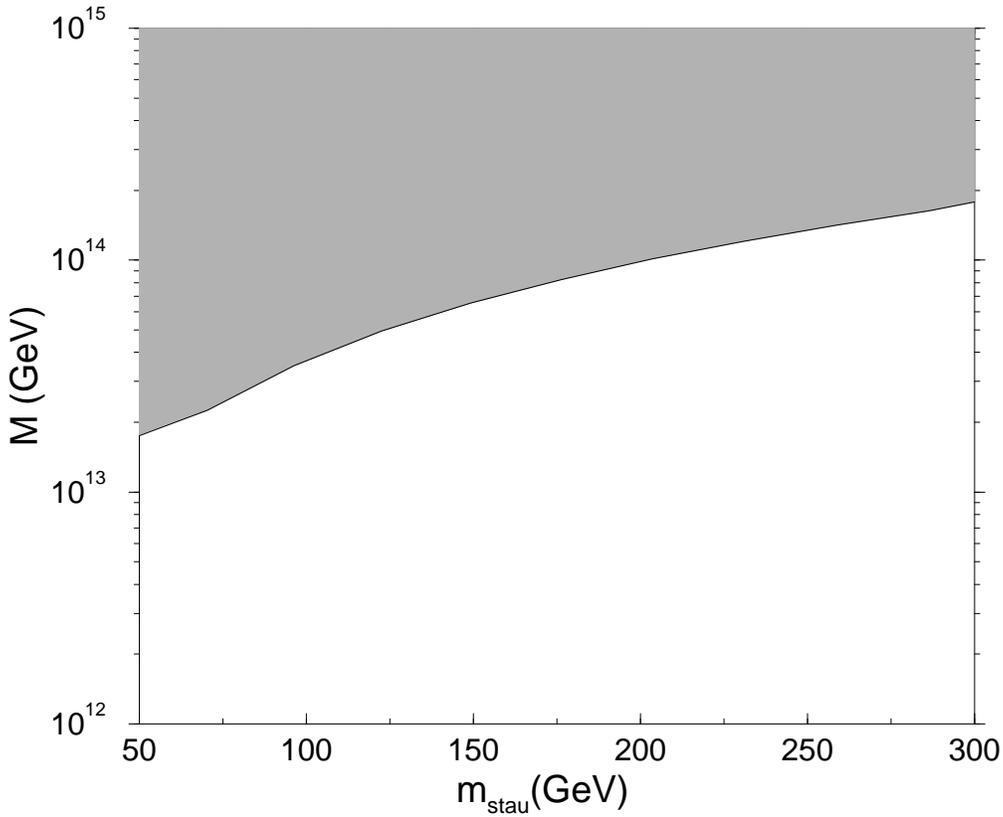}}
\caption{\it Upper bound on the messenger scale $M$ as a function
of the stau NLSP mass $m_{\tilde\tau}$ from hadronic decays
where $k=1$, $N=12$,
$\tan\beta=1.1$ and ${\rm sgn}\mu=1$.}
\label{dBstaufig}
\end{figure}

With respect to a neutralino NLSP of equal mass, 
a stau NLSP has
a larger hadronic branching ratio $B_h$, but a smaller relic abundance
because of the additional
annihilation channels. The two effects roughly compensate each other, 
although
the bound on the NLSP lifetime 
is slighly weaker in the ${\tilde \tau}_1$ case. When expressed in terms of 
$M$, the bound appears even weaker because we have to
choose a very large value of $N$ in order to satisfy the
requirement of Eq.~(\ref{limn}) for a stau NLSP. The bound on the messenger 
mass for $N=12$ and $\tan\beta =1.1$ is shown in Fig.~\ref{dBstaufig}. 
For this choice of parameters, the selectron and the smuon dominantly
decay directly into gravitinos. In the figure we only show the bound arising
from the low deuterium value. The bound for the high deuterium value 
corresponds to a lifetime which cannot be achieved for this choice of $N$.

{\bf 4.} Although the mass spectrum of supersymmetric theories
with gauge mediation can be predicted in terms of a few parameters, the
experimental and cosmological features of the theory can drastically
change as the messenger mass $M$ is varied in the allowed range between
about 100 TeV and $10^{15}$ GeV. It is thus very important to study possible
constraints on the value of $M$. The requirement that the NLSP decay products
do not upset the successful predictions of standard big-bang nucleosynthesis
provides such a constraint. Here we have studied in detail the effects of
the decaying NLSP on the light-element primordial abundances 
and shown that the most dangerous damages come from the hadronic decay modes.
After
computing the NLSP relic abundance, including coannihilation effects, we
conclude that the injection of hadronic jets in the Universe at times later
than $10^4$ seconds grossly overproduces $^7$Li. This is true for both
neutralino and stau NLSP. At earlier times, the most relevant bounds
come from deuterium overproduction. At present, there is a disagreement 
on the extracted observational value of the primordial deuterium. 
Using $X_D=(3.39\pm0.25)\times 10^{-5}$~\cite{bt}, no further limit on
the NLSP is derived, while with $X_D=(1.9\pm0.5)\times 10^{-4}$~\cite{rh}
the limit from $^7$Li can be improved. The corresponding limits on the 
messenger mass $M$ are shown in Figs.~\ref{dBneutfig} and \ref{dBstaufig}
for a representative choice of gauge mediation parameters, in the case
of neutralino and stau NLSP respectively.

It should be noticed that the bounds on the messenger mass
presented here can be evaded in the presence of other interactions which can 
lead to a fast NLSP decay, {\it e.g.} $R$-parity violation. Nevertheless
these bounds have significant implications for model building. In particular
they disfavour models in which $M$ is close to the GUT scale or models
in which the messenger scale results from balancing renormalisable interactions
with non-renormalisable operators at the Planck scale~\cite{pt,mmr}.

It is also interesting to notice that the nucleosynthesis bound discussed
here is complementary to the bound obtained from gravitino overabundance. 
Indeed when $m_{\tilde G}$ is larger than about a keV, gravitinos which were 
in thermal equilibrium at early times, give a contribution to the present
energy density larger than the critical value. It is then necessary to assume
that gravitinos have been diluted by some mechanism. Let $T_{max}$ be
the temperature at which the ordinary radiation-dominated Universe begins.
This corresponds to either the reheating temperature after an inflationary 
epoch or to the temperature associated with significant entropy production. 
The requirement that gravitinos do not overclose the Universe gives the
following contraints on $T_{max}$~\cite{mormur,gmm}
\begin{equation}
T_{max} \lappeq 100~{\rm GeV}- 1~{\rm TeV}~~~~~~{\rm for}~~~2h^2\:{\rm keV}
\lappeq m_{\tilde G}\lappeq 100~{\rm keV},
\end{equation}
\begin{equation}
T_{max} \lappeq 10~{\rm TeV}\:h^2\left( \frac{m_{\tilde G}}{100~{\rm keV}}
\right)\left( \frac{\rm TeV}{m_{\tilde g}}\right)^2~~~~~~{\rm for}~~~
m_{\tilde G}\gappeq 100~{\rm keV},
\end{equation}
where $h$ is the Hubble constant in units of 100 km sec$^{-1}$ Mpc$^{-1}$
and $m_{\tilde g}$ is the gluino mass. To compare this bound with the
nucleosynthesis result, it is convenient to express it in terms of the
messenger mass $M$, the messenger index $N$, and the pure $B$-ino mass 
$m_{N_1}$. We find
\begin{equation}
T_{max} \lappeq 100~{\rm GeV}- 1~{\rm TeV}~~~~~~{\rm for}~~~10^8~h^2~
{\rm GeV}\lappeq \frac{M}{kN}\left( \frac{m_{N_1}}{100~{\rm GeV}}\right)
\lappeq 5\times 10^9~{\rm GeV},
\end{equation}
\begin{equation}
T_{max} \lappeq 5\times 10^{-6}~h^2 \frac{M}{kN}\left( \frac{100~{\rm GeV}}
{m_{N_1}}\right)~~~~~~{\rm for}~~~\frac{M}{kN}\left( \frac{m_{N_1}}
{100~{\rm GeV}}\right) \gappeq 5\times 10^9~{\rm GeV}.
\end{equation}
This shows that there is no gravitino problem as long as\break $M\lappeq
10^8~h^2~{\rm GeV}~kN(100~{\rm GeV}/m_{N_1})$. For larger values of
$M$, there is a very stringent constraint on $T_{max}$ which requires
inflation at particularly low temperatures. As $M$ grows this limit
becomes weaker, but eventually the nucleosynthesis bound on the messenger
mass sets in. When the two bounds are combined, we find that the reheating
temperature after inflation should be less than about $10^7~h^2$ GeV
which is stronger than the bound $\sim 10^{10}$ GeV usually obtained in 
gravity-mediated scenarios\cite{dim,ell,mormur}. Our limit is valid for 
gauge-mediated theories in which the LSP gravitino is heavier
than a keV and in which there are no new interactions uncorrelated with
the supersymmetry-breaking scale mediating the NLSP decay (like $R$-parity
violating interactions). 

\bigskip

\section*{Acknowledgements}

We thank T. Moroi and C. Wagner for discussions and are especially
indebted to Sasha Pukhov for 
help in installing and running the {\tt CompHEP} software package.

\vfil\eject

\end{document}